\documentclass[superscriptaddress,aps,preprintnumbers,amsmath,showpacs,amssymb,prd,nofootinbib,reprint]{revtex4-1}
\pdfoutput=1
\usepackage{booktabs} 
\usepackage{array} 
\usepackage[table,xcdraw]{xcolor} 
\usepackage{amsmath,amssymb}
\usepackage{graphicx}
\usepackage{fancyhdr}
\usepackage{bm, color}
\usepackage{slashed,amsthm,amsfonts,empheq}
\usepackage[caption=false]{subfig}
\usepackage{hyperref}
\usepackage{booktabs}
\usepackage{multirow}
\usepackage[left=2cm,right=2cm,top=2cm,bottom=2cm,includefoot,a4paper]{geometry}
\usepackage{tikz}
\usetikzlibrary{arrows.meta}

\newcommand{\Slash}[1]{{\ooalign{\hfil#1\hfil\crcr\raise.167ex\hbox{/}}}}

\newcommand{\beq}{\begin{equation}}  \newcommand{\eeq}{\end{equation}}
\newcommand{\bef}{\begin{figure}}  \newcommand{\eef}{\end{figure}}
\newcommand{\bec}{\begin{center}}  \newcommand{\eec}{\end{center}}

\newcommand{\Sec}[1]{Sec.\ref{chap:#1}}

\newcommand{\vev}[1]{\left\langle {#1} \right\rangle}

\newcommand{\lac}[1]{\label{chap:#1}}
\newcommand{\SU}[1]{{\rm SU{#1} } }

\def\({\left(}
\def\){\right)}

\def\O{\mathcal{O}}

\def\tr{\mathop{\rm tr}}

\def\a{\alpha}

\def\d{\delta}

\def\k{\kappa}

\def\x{\xi}

\def\L{\Lambda}

\def\*{\dagger}

\newcommand{\cM}{\mathcal M}

\def\WY#1{#1}

\begin{document}

\title{Why 4D? Spontaneous Dimensional Selection from Gauge Criticality}

\author{Wen Yin}
\affiliation{Department of Physics, Tokyo Metropolitan University, Minami-Osawa, Hachioji-shi, Tokyo 192-0397, Japan}

\begin{abstract}
Noting that the Yang--Mills coupling is relevant below four dimensions,
marginal in four dimensions, and irrelevant above four dimensions, I propose
{that this criticality can lead to the selection of four macroscopic dimensions.
As a proof of principle, I present a concrete racetrack model of the type
commonly studied in extra-dimensional scenarios, but without assuming a
noncompact 4D spacetime.  I find, within a well-controlled model without specifying the 4D, that 4D
spacetime is spontaneously realized through radion stabilization via
gauge-criticality-induced supersymmetry breaking and supergravity effects.}
\end{abstract}

\maketitle

\section{Introduction}

Why does the observed Universe have four macroscopic spacetime dimensions (4D)?
The question is particularly sharp because candidate fundamental theories
often admit many higher-dimensional compactification patterns.  Consistent
quantum field theories (QFTs) also exist in
lower dimensions: for example, $(2+1)$-dimensional theories routinely describe
low-energy condensed-matter systems.  

I approach the question from the perspective of higher-dimensional QFT, focusing on the
critical dimensionality of gauge interactions. For a pure non-Abelian gauge sector, the leading local
operator is the Yang--Mills kinetic term.  Its coupling obeys
\begin{equation}
 [g_D^2]=4-D.
 \label{eq:critical}
\end{equation}
Thus it is relevant below four dimensions, marginal in four dimensions, and
irrelevant above four dimensions.  This suggests that low-energy confinement
dynamics can strongly affect spacetime moduli, such as the radii, for
$D\leq4$, while becoming irrelevant for $D>4$. {I therefore consider two possible scenarios}
leading to a 4D universe:
\begin{itemize}
\setlength{\topsep}{2pt}
\setlength{\itemsep}{0pt}
\setlength{\parsep}{0pt}
\setlength{\parskip}{0pt}
\item[(a)]{The parent theory does not assume a noncompact 4D spacetime;
a 4D regime is dynamically selected by gauge criticality.}
\item[(b)] Several vacua of different dimensionalities exist, but only the 4D gauge phase can survive
 because of the gauge-induced contribution and the associated cosmology (see \Sec{discussion}).
\end{itemize}

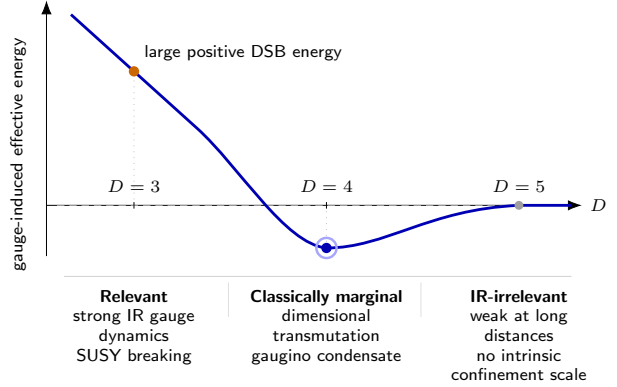
\begin{figure}[t]
\centering
\resizebox{\columnwidth}{!}{%
\begin{tikzpicture}[
  x=1cm,y=0.58cm,
  every node/.style={font=\sffamily\scriptsize},
  regime/.style={align=center,anchor=north,text width=2.25cm}
]
 \draw[-{Latex[length=2mm]}] (0,0) -- (7.65,0)
   node[right] {$D$};
 \draw[-{Latex[length=2mm]}] (0,-1.25) -- (0,5.05);
 \node[rotate=90,anchor=south] at (-0.18,1.25)
   {gauge-induced effective energy};
 \draw[gray!65,densely dashed] (0.15,0) -- (7.50,0);
 \draw[gray!45,dotted] (1.25,0) -- (1.25,3.30);
 \draw[gray!45,dotted] (4.00,0) -- (4.00,-1.05);
 \draw[blue!70!black,line width=1.15pt]
   (0.35,4.70)
   -- (2.15,1.90)
   .. controls (2.75,0.97) and (3.40,-0.95) ..
   (4.00,-1.05)
   .. controls (4.85,-1.05) and (5.45,-0.10) ..
   (6.75,0)
   -- (7.50,0);
 \foreach \x/\d in {1.25/3,4.00/4,6.75/5}{
   \draw (\x,0.08) -- (\x,-0.08);
   \node[above=2pt] at (\x,0) {$D=\d$};
 }
 \fill[orange!80!black] (1.25,3.30) circle (2.1pt);
 \fill[blue!70!black] (4.00,-1.05) circle (2.1pt);
 \draw[blue!35,line width=1pt] (4.00,-1.05) circle (4.1pt);
 \fill[gray!75] (6.75,0) circle (1.8pt);
 \node[above right=1pt] at (1.25,3.30) {large positive DSB energy};
 \draw[gray!35] (0.25,-1.75) -- (7.50,-1.75);
 \draw[gray!25] (2.60,-1.90) -- (2.60,-2.75);
 \draw[gray!25] (5.35,-1.90) -- (5.35,-2.75);
 \node[regime] at (1.25,-1.88)
   {\textbf{Relevant}
   \\strong IR gauge dynamics\\SUSY breaking};
 \node[regime] at (4.00,-1.88)
   {\textbf{Classically marginal}\\dimensional transmutation\\gaugino condensate};
 \node[regime] at (6.75,-1.88)
   {\textbf{IR-irrelevant}\\weak at long distances\\no intrinsic confinement scale};
\end{tikzpicture}
}
\caption{Schematic dimensional dependence of the gauge-induced contribution
to the matched radion potential.  The vertical scale is qualitative.}
\label{fig:dimensionalEnergy}
\end{figure}

 In this paper, I mainly study {scenario} (a). As a proof of principle,
{I employ an extra-dimensional pure supersymmetric Yang--Mills (SYM) model.}
In this model, nonperturbative gauge dynamics is suppressed in the 5D effective
field theory (EFT), the radion can be stabilized in 4D by gauge confinement,
{and supersymmetry (SUSY) is spontaneously broken after compactification to 3D, generating a}
large positive vacuum energy.
Without an uplift to de Sitter space, the 4D AdS vacuum has negative energy and
is therefore the lowest.  The qualitative pattern is summarized in
Fig.~\ref{fig:dimensionalEnergy}.
{Therefore, if the 4D vacuum is uplifted to de~Sitter space while the
hierarchy between the macroscopic and microscopic radii is preserved or enhanced, the
result is a 4D Universe similar to ours.}

Before going into detail, let me mention existing proposals addressing the
``why 4D?'' question. They can be divided into two broad groups:
\begin{itemize}
\setlength{\topsep}{2pt}
\setlength{\itemsep}{0pt}
\setlength{\parsep}{0pt}
\setlength{\parskip}{0pt}
 \item \emph{{Anthropic and consistency arguments at fixed dimensionality.}}
{The stability of the Weyl equation and of bound systems}, predictability of time
evolution, and asymptotic safety have each been used to favor four dimensions
\cite{Nielsen:1993fd,Tegmark:1997jg,Eichhorn:2019yzm}.
\item \emph{{Dynamical selection or emergence of dimensionality.}}
{Cosmological mechanisms such as string-gas winding annihilation and
confining flux-tube networks can select three large spatial dimensions
\cite{Brandenberger:1988aj,Greene:2009gp,Berera:2015yna}, while a Gaussian
expansion of the Euclidean IIB matrix model favors an $SO(4)$-symmetric
emergent geometry among those studied \cite{Nishimura:2001sx}.}
\end{itemize}
This work belongs to the second group.

 Nonperturbative gauge
 superpotentials are standard tools for radion stabilization
\cite{Krasnikov:1987jj,Kounnas:1988ye,Casas:1990qi, deCarlos:1992kox, Kachru:2003aw,Kallosh:2004yh}; there, however, 4D spacetime is assumed to be noncompact, {thereby singling it out}. 
In the concrete model of this paper, {the radius associated with a lower-dimensional EFT} is also treated as dynamical. {In principle, this radius could be stabilized at a value as small as the other compact radius, but I show in a concrete model that gauge dynamics makes the former much larger than the latter.}

\section{A supersymmetric parent model}

\subsection{Geometry and symmetries}

I consider a supergravity (SUGRA) theory with a bulk vector multiplet on
\begin{align}
 \cM_5&=\mathbb R^{2,1}\times I_4\times I_5,\nonumber\\
 I_i&=S^1/\mathbb Z_2,\qquad L_i=\pi R_i,\qquad i=4,5.
 \label{eq:geometry}
\end{align}
Both radii $L_i$ are dynamical.  In SUGRA, the bosonic and fermionic
contributions of complete KK multiplets cancel, suppressing the radiative
radion potential.  Thus I employ SUSY for the proof of principle, because this radiative correction and can make the gauge contribution dominant in radii stabilization. 

{I keep the other two spatial directions noncompact throughout the explicit
construction.  Promoting them to compact dynamical directions, denoted by
$I_{2,3}$, is model dependent because their manifold, boundary
conditions, and finite-volume dynamics must be specified.  In particular,
for $I_{2,3}=(S^1/\mathbb Z_2)^2$, the fixed strata admit localized
tensions and other operators whose coefficients are independent matching
data and are not fixed or na\"{i}vely protected by SUSY.  The corresponding
radion potential is therefore not determined by the bulk theory alone. 
I do not include $I_{2,3}$ as dynamical compact dimensions in this work for simplicity.}
Assuming the ordering
$L_4>L_5$,\footnote{At $L_4\simeq L_5$, the intermediate 4D Wilsonian window
closes, but the crossing is nonsingular.  The appropriate description is the
3D minimal-SUSY EFT obtained by reducing on $I_4\times I_5$ while retaining
both radions.  Its leading bulk coupling satisfies
{$1/g_{3,r}^2=L_4L_5/g_{5,r}^2$}, so the positive DSB energy depends on the product
 $L_4L_5$ and increases when either radius decreases.  It therefore does
 not produce a stationary point at $L_4\simeq L_5$; increasing either radius
 lowers this contribution{, although the resulting 3D potential is still higher than the AdS potential}.} the hierarchy $L_4>\mu_{\rm RG}^{-1}>L_5$ defines a four-dimensional
Wilsonian EFT at the scale $\mu_{\rm RG}$.  The question is whether the radion can be
stabilized in a regime with $L_4\gg L_5$ without imposing a setup that
specifies 4D in advance.

I also assume the following properties of the parent theory, none of which
singles out 4D:
\begin{itemize}
\item The EFT in each effective dimension has the minimal SUSY appropriate to
that dimension.  This can be realized using orbifold projections (see
Appendix~\ref{app:2} for details).
\item {In the 5D parent theory I impose time reversal $\mathsf T_5$ and
an appropriate, possibly discrete, $R$ symmetry $\mathsf R$.
I choose the orbifold projections so that the anti-unitary combination}
\begin{equation}
{
 \mathsf T_{\rm IR}\equiv\mathsf T_5\mathsf R
}
\label{eq:IRtimeReversal}
\end{equation}
is preserved in the 3D IR EFT and acts as time reversal.  It forbids the dangerous 3D superpotential term in the corner, which is otherwise not protected by any symmetry.\footnote{Without this assumption, arbitrary constant and
radion-dependent terms can be placed on the 3D fixed strata; the resulting
theory is uncontrolled.} It also forbids a
nonzero Chern--Simons level. For
the adjoint Majorana gaugino,  $k_{\rm CS}-h^\vee/2\in\mathbb Z$ is needed; I therefore
take even $N$ for each $\SU(N)$ sector, so that $k_{\rm CS}=0$ is
allowed~\cite{Witten:1999ds}.   At $k_{\rm CS}=0$, the 3D theory is believed
{to break SUSY dynamically, as predicted in
Refs.~\cite{Witten:1999ds,Gomis:2017ixy}.}\footnote{Even
if SUSY remains unbroken in 3D, the 4D racetrack AdS minimum in the $T$
direction persists, and a conventional uplift can turn it into a local
Minkowski minimum in the same direction.  Thus 3D DSB is not required for
$L_5$ stabilization but rather a plausible prediction.} 
\end{itemize}

\subsection{Four-dimensional vacuum}

{I first study} the system using an EFT at a renormalization scale appropriate to
a 4D description.  This choice merely sets the energy scale of interest and
does not specify 4D by hand.
The advantage of using the 4D EFT is that it permits the use of the familiar framework of QFT.

Let $T$ denote the chiral radion of $I_5$. Assuming multiple pure SYMs of $\SU(N_r)$ in the bulk, the tree-level 4D SUSY EFT
{is fixed by the symmetries as}
\begin{align}
{\cal L}_{\rm gauge}
 &=\sum_{r}\int d^2\theta\,
 f_r\tr{\cal W}_r^\a{\cal W}_{r\a},
 \label{eq:seq2}\\
{\cal K}_4
 &=-3\log(T+T^\*),\qquad W_4=0,
 \nonumber\\
f_r
 &=\frac{T}{2g^2_{5,r}}
 +\frac{3N_r}{16\pi^2}\log\left(\frac{\mu_{\rm RG}}{\L_{5,r}}\right).
 \label{eq:gauginoNew}
\end{align}
{I use four-dimensional Planck units.} Notice that the imposed $R$-symmetry forbids the constant superpotential.
Here $\Re T=L_5$, while its imaginary part is associated with the
graviphoton.  The coefficient $3$ in ${\cal K}_4$ follows from dimensional reduction of the
5D gravitational action.  The radion satisfies
\beq
T\to T+i \a
\eeq
where $\a$ is real.  This follows from the perturbative graviphoton shift
symmetry, whose discrete subgroup is exact.
{I also neglect higher-order K\"ahler corrections induced by the Casimir energy of the
extra-dimensional KK modes.}
{More generally, I assume that gravitational-loop corrections and
nonzero-KK-mode loop corrections to the low-energy EFT are parametrically
small and neglect them, except for the gauge-threshold matching included
explicitly below.}
Independently, summing the perturbative gauge threshold effects gives
{for each sector}~\cite{Luty:1999cz}
\beq
\L_{5,r}= \frac{\k_r}{g^{2}_{5,r}},
\eeq
which can be also understood from holomorphy and the shift symmetry. 
Here $\L_{5,r}$ is a radion-independent 5D matching scale, not a
confinement scale, and $\k_r$ is a dimensionless radion-independent constant.
Radion-independent fixed-point
 gauge kinetic terms are absorbed into $\k_r$ and hence into $\L_{5,r}$.

{I assume the natural cutoff-scale estimates}
\begin{equation}
 \frac{1}{g_{5,r}^2}\sim\frac{N_rM_5}{24\pi^3},
 \label{eq:NDAcoupling}
\end{equation}
where $M_5$ is the 5D Planck scale and also the cutoff scale of
the full parent theory. 
The two nonperturbative sectors are taken to have
order-one ratios of their exponents and prefactors; a common hierarchy
between their dynamical scales and $M_5$ is allowed.

{Each pure-SYM sector in infinite 4D volume confines at}
\begin{align}
 \L_r^3(T)&=|A_r|e^{-a_r\Re T},
~~~ a_r=\frac{8\pi^2}{N_rg_{5,r}^2},
 \label{eq:twoConfinementScales}
\end{align}
The relevant regimes are distinguished by the multiple confinement
thresholds $\L_r(L_5)$.
At fixed $(L_4,L_5)$, define
$\mathcal S_4=\{r\,|\,L_4^{-1}<\L_r\}$ and
$\mathcal S_3=\{r\,|\,L_4^{-1}>\L_r\}$.  A sector in $\mathcal S_4$
confines in four dimensions, whereas a sector in $\mathcal S_3$ is first
matched to the three-dimensional EFT.  These descriptions are controlled
away from the individual threshold bands $L_4\L_r=\O(1)$.

{To discuss the two cases $\mathcal S_3=\varnothing$ and
$\mathcal S_3\neq\varnothing$,} I treat $L_4$ as a background and study
the dynamics of $T$ within the regime of validity of each EFT and then
{determine} the vacuum structure of the system. {Equivalently, after
integrating over the compact directions, the potential is defined by the
Euclidean partition function as}
{
\begin{equation}
 V_3^E(L_4,L_5)
 =-\lim_{\mathcal V_{3,E}\to\infty}
 \frac{1}{\mathcal V_{3,E}}\log Z_E(L_4,L_5),
 \label{eq:partitionPotential}
\end{equation}
}
{where $\mathcal V_{3,E}$ is the volume of the common noncompact
Euclidean spacetime.}
\paragraph{{Case 1: $\mathcal S_3=\varnothing$}}

From now on, for simplicity, I consider only two SYM sectors in the bulk.\footnote{With further SYMs, the strongest two would balance each other for the stabilization of $L_5$, while 
 $L_4$ would obtain larger vacuum expectation value due to the DSB for the weaker SYMs.}
{Below I write $(A_1,A_2)=(A,B)$ and $(a_1,a_2)=(a,b)$.}

A 4D description of gaugino condensation is then valid, and the resulting 4D
{potential can stabilize $L_5$ through the two-condensate racetrack~\cite{Krasnikov:1987jj,Kounnas:1988ye,Casas:1990qi, deCarlos:1992kox,Kallosh:2004yh}
superpotential}
\begin{equation}
 W_{\rm rt}=Ae^{-aT}+Be^{-bT},
 \qquad A>0,\quad B<0,\quad a,b>0 .
 \label{eq:racetrackW}
\end{equation}
{The} 
relevant SUGRA potential is
\begin{align}
 V_{4D}
 &=e^{{\cal K}_4}\left(
 K^{T\bar T}|D_T\WY{W_{\rm rt}}|^2-3|\WY{W_{\rm rt}}|^2\right).
 \label{eq:Gamma4}
\end{align}
{The supersymmetric stationary condition is $D_TW_{\rm rt}=0$, where
$D_TW_{\rm rt}\equiv\partial_TW_{\rm rt}
+(\partial_T{\cal K}_4)W_{\rm rt}$.  At a stationary point
$T=t_*\in\mathbb R$, it is equivalent to}
\begin{equation}
 \frac{B}{A}
 =-e^{(b-a)t_*}
 \frac{a+3/(2t_*)}{b+3/(2t_*)}.
 \label{eq:minimumNew}
\end{equation}
{At this stationary point, Eq.~\eqref{eq:minimumNew} also gives
\begin{equation}
 \frac{\L_2}{\L_1}
 =\left(\frac{a+3/(2t_*)}{b+3/(2t_*)}\right)^{1/3}.
 \label{eq:scaleRatioAtMinimum}
\end{equation}
Thus order-one racetrack parameters make the two confinement scales
parametrically of the same order, although not exactly equal.}
Then
\beq 
V_{4D}=-\vev{3e^{{\cal K}_4}
|\WY{W_{\rm rt}}|^2}<0.
\eeq 

{The AdS vacuum can be uplifted by SUSY-breaking.}
Here I retain the resulting AdS vacuum and do not introduce an uplift; this
possibility is discussed together with cosmology in \Sec{discussion}.

{Both gauge sectors confine before}
compactification: the
gapped Yang--Mills degrees of freedom have already been integrated out.
{The low-energy EFT obtained by compactifying the 4D theory contains
gravity, the radion, and the inherited cosmological term.}  Define
\begin{equation}
 \rho\equiv\log\frac{L_4}{L_{4,0}},\qquad
 ds_4^2=e^{-2\rho}g^E_{\mu\nu}dx^\mu dx^\nu
       +e^{2\rho}dy_4^2,
 \label{eq:4Dto3Dmetric}
\end{equation}
where the coordinate interval has reference length $L_{4,0}$.  In 3D Planck
units the scalar kinetic terms take the form
\begin{equation}
 e^{-1}{\cal L}_{3,\mathrm{kin}}
 =\frac12{\cal R}_3-(\partial\rho)^2
  -K_{T\bar T}\partial T\partial\bar T+\cdots .
 \label{eq:3DtwoFieldKinetic}
\end{equation}
Thus the real scalar coordinates are
$q^A=(\rho,t,a)$, with $T=t+ia$ and $t=L_5$.  A two-derivative 3D minimal-SUGRA
potential can be written as~\cite{Emelin:2021gzx}
\begin{equation}
 V_{3,\mathrm{SUGRA}}
 =G^{AB}\partial_A{\cal P}_3\partial_B{\cal P}_3
  -4{\cal P}_3^2,
 \label{eq:3DSUGRApotential}
\end{equation}
where ${\cal P}_3$ is a real superpotential.

For a 4D chiral sector {after gaugino condensation}, introduce the K\"ahler-invariant real function
\begin{equation}
{{\cal W}_{\rm rt}(T,\bar T)\equiv
 e^{{\cal K}_4/2}|W_{\rm rt}|.}
 \label{eq:realW4}
\end{equation}
The 4D F-term potential obeys the identity
\begin{equation}
 V_{4D}
 =4K^{T\bar T}\partial_T{{\cal W}_{\rm rt}}
                 \partial_{\bar T}{{\cal W}_{\rm rt}}
  -3{{\cal W}_{\rm rt}^2}.
 \label{eq:V4realW}
\end{equation}
Up to the overall normalization fixed by
$M_3=M_4^2L_{4,0}$, the dimensional reduction is reproduced by
\begin{equation}
{
 {\cal P}_3=e^{-\rho}{{\cal W}_{\rm rt}} .
}
 \label{eq:P3fromW4}
\end{equation}
Indeed, the $\rho$ derivative contributes $+{\cal P}_3^2$, the two real
components of $T$ reproduce the first term in Eq.~\eqref{eq:V4realW}, and the
last term in Eq.~\eqref{eq:3DSUGRApotential} contributes
$-4{\cal P}_3^2$.  Restoring the reference length gives
\begin{equation}
 V^E_{3,\mathrm{SUGRA}}(L_4,T)
 =\frac{L_{4,0}^3}{L_4^2}V_{4D}(T).
 \label{eq:3Dfrom4D}
\end{equation}
{This can also be obtained by directly compactifying the fourth dimension using $V_{4D}$.}

\paragraph{{Case 2: $\mathcal S_3\neq\varnothing$}}
{Every sector is matched separately.  For $r\in\mathcal S_4$, the
four-dimensional condensate $W_r=A_re^{-a_rT}$ is retained.  For
$r\in\mathcal S_3$, the gauge theory is reduced at $\mu_{\rm RG}=L_4^{-1}$ and gives
a three-dimensional DSB contribution.  Thus Case 2 includes both mixed
regimes and the regime in which all sectors are three-dimensional.}  The
orbifold projection removes the scalar zero mode from the extra component of
{the gauge fields, so the surviving sectors are 3D minimal-SUSY pure
SYM theories.}
{Time reversal sets $k_{{\rm CS},r}=0$ for every
$r\in\mathcal S_3$.  At these levels the corresponding theories are believed
to break SUSY dynamically}~\cite{Witten:1999ds,Gomis:2017ixy}.
{After integrating out each DSB gauge sector below its mass gap, its
leading contribution is}
\begin{equation}
 V_{3,\mathrm{DSB}}={\sum_{r\in\mathcal S_3}
 \xi_{3,r}\left(N_rg_{3,r}^2\right)^3},
 \qquad {\xi_{3,r}>0}.
 \label{eq:3DDSB}
\end{equation}
{Here the positive sign follows from spontaneous SUSY breaking, and
$\xi_{3,r}$ is expected to be $\O(1)$.}  Reducing the
 gauge kinetic term gives {for each sector}
\begin{equation}
 \frac{1}{{g_{3,r}^2}}
 =\frac{L_{4}}{\WY{g_{4,r}^2(L_4^{-1})}},
 \qquad
{g_{4,r}^2(L_5^{-1})=\frac{g_{5,r}^2}{L_5}}.
 \label{eq:3Dcoupling}
\end{equation}
\WY{Here $g_{4,r}^2(L_4^{-1})$ is obtained by RG evolution from the 5D--4D
matching condition at $L_5^{-1}$.}
\WY{Thus the leading matched DSB energy in the metric inherited from the
4D theory is}
\begin{equation}
 V_{3,\mathrm{DSB}}(L_4,L_5)
 =\WY{\sum_{r\in\mathcal S_3}\xi_{3,r}
 \left(\frac{N_rg_{4,r}^2(L_4^{-1})}{L_4}\right)^3}.
 \label{eq:3DDSBradii}
\end{equation}
Introducing a fixed reference length $L_{4,0}$, the Weyl transformation
gives \WY{its contribution to} the Einstein frame potential
\begin{equation}
 V^{E}_{3,\mathrm{DSB}}
 =\WY{\sum_{r\in\mathcal S_3}
 \xi_{3,r}\!\left[N_rg_{4,r}^2(L_4^{-1})\right]^3}
 \frac{L_{4,0}^3}{L_4^6}.
 \label{eq:3DDSBEinstein}
\end{equation}
\WY{The matched Einstein-frame DSB contributions in
Eq.~\eqref{eq:3DDSBEinstein} are added sector by sector to the descended
condensate potential.}

\begin{figure}[t]
\centering
\includegraphics[width=\columnwidth]{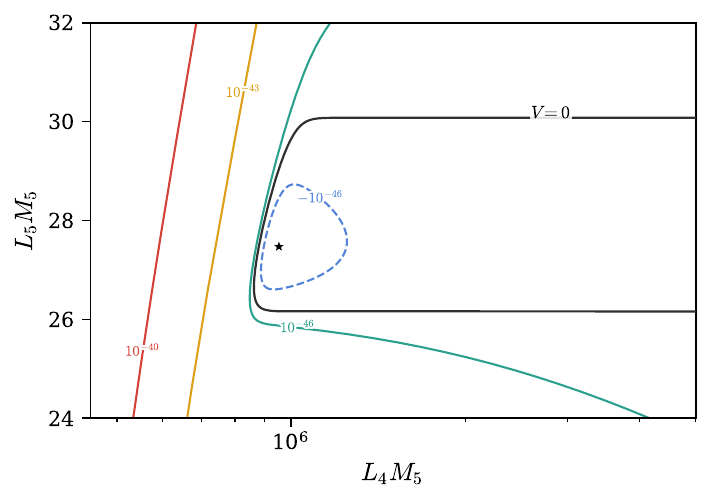}
\caption{\WY{Contours of $V_{3,\mathrm{eff}}/M_{5}^3$.  The
star marks the AdS minimum,
\WY{$(L_4M_5,L_5M_5)\simeq(9.52\times10^5,27.47)$, where
$V_3/M_5^3\simeq-1.63\times10^{-46}$.}
For $L_4\cdot L_5\to \infty$ the potential approaches zero.  }}
\label{fig:2}
\end{figure}

\paragraph{Vacuum structure}
\WY{The threshold bands $L_4\L_r=\mathcal O(1)$ are nonperturbative.
For illustration, I connect the separately matched sectors with smooth
sectorwise switches.}
For the contour plot of the potential in Fig.~\ref{fig:2}, I use
\begingroup
\begin{align}
h_r&=1-s_r,\qquad
s_r(L_4,L_5)=e^{-L_4\Lambda_r(L_5)},
\nonumber\\
{\cal P}_{3,\mathrm{conf}}
&=e^{-\rho}e^{{\cal K}_4/2}
\left|\sum_r h_rW_r\right|,
\nonumber\\
V_{3,\mathrm{eff}}^E
&=G^{AB}\partial_A{\cal P}_{3,\mathrm{conf}}
\partial_B{\cal P}_{3,\mathrm{conf}}
-4{\cal P}_{3,\mathrm{conf}}^2
\nonumber\\
&\quad
+\frac{L_{4,0}^3}{L_4^6}
\sum_r s_r\,\xi_{3,r}
\left[N_rg_{4,r}^2(L_4^{-1})\right]^3.
\label{eq:sectorSwitches}
\end{align}
\endgroup
\WY{The derivatives in this expression also act on $h_r$.  }
For
$N_1=N_2=2$ and $\xi_{3,1}=\xi_{3,2}=1$, I use
\WY{$(a/M_5,b/M_5)=(0.10,0.15)$ and
$(A/M_5^3,B/M_5^3)=(10^{-12},-3\times10^{-12})$. }
\WY{Numerical minimization gives
$L_4/L_5\simeq3.47\times10^4$ at the potential minimum.} Therefore a large hierarchy of the potential is obtained.\footnote{With $a\approx b$ a even larger hierarchy can be obtained. The choice of the parameter is for visual of the figure. } 

\WY{Although I use a smooth transition between the DSB and condensate
descriptions, which may be preferred in the $N=2$ case, e.g., a first-order transition can change its local details
but not the controlled asymptotics: the potential
approaches zero as $L_4\cdot L_5\to\infty$, is negative on the confined branch. It is positive on the DSB branch.  Thus vacuum selection generically
favors $L_4 \gg L_5$ which is around the boundary of the two branch.}

\section{Conclusions and Discussion}
\lac{discussion}

I have shown that the 4D AdS spacetime can be realized through
the criticality of gauge interactions and the relevance of nonperturbative
effects.  Although the concrete model only serves as a proof of principle, the emergence of 4D through gauge
criticality should also be possible in more generic gauge-theory models.

\paragraph{Uplift of the AdS vacuum}
It is well known that our current Universe is not described by the AdS
vacuum, and the natural question is whether the 4D selection mechanism is consistent with
our Universe.
{Although I use SUSY as a proof of principle,} compatibility
with Minkowski or de~Sitter space is a separate issue.  The vacuum can be
{uplifted, e.g., with a nilpotent chiral multiplet $X$ in the 4D description:
$X^2=0$ and $R[X]=2$.  Taking $\mu^2$ real by a phase redefinition,
$K=K_4+X\bar X$ and
$W=W_{\rm rt}+\mu^2X$ generate a positive contribution. $\mu^2$ can be small by an approximate discrete symmetry of $X$.}
{
{Several other uplift mechanisms are possible, including ones originating in
the 5D parent theory~\cite{Abe:2007zv}. }
Along the finite-$L_5$
branch, the uplifted potential still selects $L_4\gg L_5$: $L_5$ remains
\WY{stabilized while $L_4$ approaches the 4D Minkowski valley
(see Appendix~\ref{app:upliftContours}).}
{The gauge-induced DSB prevents $L_4$ from becoming microscopic.
}

In this minimal
uplift, $L_4$ is not fixed at a finite value, and the simultaneous 5D
decompactification limit remains asymptotically degenerate.
Whether the cosmological evolution reaches the 4D minimum therefore depends on the
initial conditions and early cosmological history.  Cosmological
trapping is also possible in the racetrack potential.  A radiation or
matter background increases Hubble friction,
slows the volume modulus, and can substantially enlarge the set of initial
conditions that approach the finite-radius minimum instead of overshooting
its barrier~\cite{Brustein:2004jp,Barreiro:2005ua}.  Thus, if the early
cosmological trajectory enters this basin, it can remain in a finite-radius
4D phase and approach the late-time 4D vacuum as the Universe expands.
Because of the runaway, these effects may drive $L_4$ to a very large
value.

{The point is that the existence of a 4D basin with stabilized $L_5$ and
$L_4\gg L_5$, obtained without specifying 4D at the outset, should be
sufficient for my purpose.}
}

\paragraph{Inflation}

{In the 3D Einstein-frame description, positive 4D vacuum energy drives
$L_4$ to larger values.  Once the system enters the regime}
$L_4^{-1}<H<L_5^{-1}$, where $H$ is the Hubble scale, the appropriate
Wilsonian description at the scale $H$ is the 4D EFT.  The radius $L_4$ is
then the third spatial scale factor, and the solution describes ordinary 4D
inflation~(e.g. \cite{Graham:2010hh, Blanco-Pillado:2010rbj}).
\WY{For instance, $\operatorname{Im}T$, whose perturbative shift symmetry is
broken by $W_{\rm rt}$, is a possible inflaton candidate.}
 A small-field natural
inflation scenario can be obtained if the potential energy during inflation
destabilizes other moduli; in this scenario, the inflaton itself may even be
a dark matter candidate~\cite{Murase:2025uwv}.  Such a radion-induced change
in the gauge coupling can be important for setting the initial condition for
QCD axion dark matter~\cite{Kitano:2021fdl,Kitano:2023mra,Dvali:2026ceb,Choi:2026tup}, if the strong CP problem is induced due to a spontaneous breaking CP for having the violation. 

\paragraph{Scenario (b)}
{Finally, I mention Scenario $(b)$, motivated by the fact that
non-SUSY pure Yang--Mills theory may make a negative contribution to the
vacuum energy, relative to a perturbative reference vacuum, through
confinement.
{Related uses of negative or positive vacuum energy are discussed in}
Refs.~\cite{Takahashi:2008pu,Bloch:2019bvc,Takahashi:2021tff,TitoDAgnolo:2021nhd}.}

{As in the explicit construction, I keep two spatial directions
noncompact, so every branch considered here has $d\geq3$ macroscopic
spacetime dimensions.  For each branch, I integrate out only its microscopic
stabilized radii and denote the resulting $d$-dimensional Einstein-frame
vacuum energy by $\rho_d$.  Vacuum energies in different dimensions are not
compared numerically; only their signs are used.  In lower-dimensional
branches,
{a dimensionful gauge coupling can enhance confinement and give}
a negative contribution to the vacuum energy.  For $d>4$, Yang--Mills
interactions are irrelevant, and
{the loss of trapping by confinement can}
instead release a scalar toward its negative-energy vacuum.  Thus}
\begin{equation}
{\rho_4\geq0,\qquad \rho_d<0\quad(d\ne4).}
 \label{eq:scenarioBsigns}
\end{equation}
{For a spatially flat, isotropic branch,}
\begin{equation}
{\frac{(d-1)(d-2)}{2}M_d^{d-2}H_d^2
 =\rho_{m,d}+\rho_d.}
 \label{eq:scenarioBfriedmann}
\end{equation}
{A negative-energy branch that initially expands therefore turns around
and recollapses after matter and radiation with the energy density $\rho_{m,d}$ dilute, whereas the 4D branch has
no vacuum-energy-induced turnaround.  Thus 4D is selected among
nonrecollapsing expanding branches.}

{The vanishingly small vacuum energy may be relaxed during inflation by a
peculiar inflaton potential~\cite{Yin:2021uus}, which predicts a time-varying
vacuum energy consistent with that inferred from
DESI~\cite{Yin:2024hba}.}

\section*{Acknowledgements}
{W. Y. thanks the organizers of the workshop ``Axion in Seoul'' for the kind hospitality, where this work was finalized.}
This work is supported by JSPS KAKENHI Grant Nos. 22K14029 (W.Y.),
23K22486 (W.Y.), and 26K00695 (W.Y.).  W.Y. is also supported by the Selective
Research Fund and the Incentive Research Fund of Tokyo Metropolitan University.

\clearpage

\appendix

\section{{Orbifold projection in the 5D model}}
\label{app:susyProjection}
\label{app:2}
{To avoid specifying 4D as a special dimension, I choose the orbifold
projections so that each EFT in the sequence has the
dimension-appropriate minimal SUSY.}
Then the 5D parent theory has minimal 5D SUSY, with eight real supercharges.  Let $r_i$ denote
reflection of $I_i$ for $i=4,5$, and let $\widehat r_i$ be its Pin lift on
the 5D spinor.  Choose two anticommuting matrices $\rho_i$ acting on the
parent $SU(2)_R$ index and define
\begin{equation}
{
 \mathsf P_i=\widehat r_i\otimes\rho_i,\qquad
 \mathsf P_i^2=1,\qquad [\mathsf P_4,\mathsf P_5]=0 .
}
\label{eq:5DcommutingParities}
\end{equation}
{The last relation follows because both the Pin lifts and the $\rho_i$
anticommute.  The orbifold conditions and surviving supercharges are}
\begin{equation}
{
\begin{aligned}
 \epsilon(r_i y_i)&=\mathsf P_i\epsilon(y_i),&
 \Pi_i&=\frac{1+\mathsf P_i}{2},\\
 {\cal Q}_{4D}&=\Pi_5{\cal Q}_{5D},&
 {\cal Q}_{3D}&=\Pi_4\Pi_5{\cal Q}_{5D}.
\end{aligned}
}
\label{eq:5DsurvivingQ}
\end{equation}
{Thus the two projections leave respectively four and two real
supercharges, giving 4D $\mathcal N=1$ and 3D $\mathcal N=1$.  For the bulk
vector, $A_\mu$ tangent to a fixed locus is even and the normal component
$A_i$ is odd; the gaugino transforms with $\mathsf P_i$.  A field must be
even under each reflection separately.  This establishes the
$5D\to4D\to3D$ construction used here.}

{The two 5D transformations need not commute with the orbifold projectors
separately.  The projections are chosen so that
$[\mathsf T_5\mathsf R,\mathsf P_i]=0$ for $i=4,5$.  Hence the combination
$\mathsf T_{\rm IR}=\mathsf T_5\mathsf R$ acts on the projected 3D
supercharges as time reversal and enforces $k_{\rm CS}=0$.  The inherited
$R$ selection rule forbids an independent constant superpotential in 4D.}

\section{{Representative Minkowski contours}}
\label{app:upliftContours}

\begin{figure}[t]
\centering
\includegraphics[width=\columnwidth]{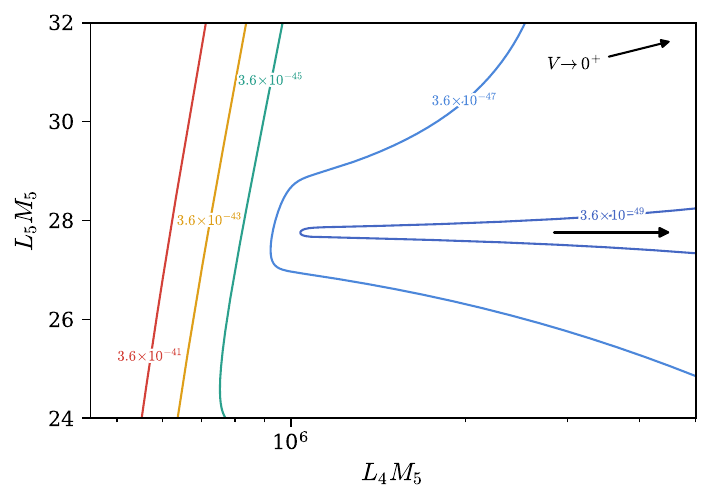}
\caption{\WY{Contours of $V_{\rm 3D}$ after the $R[X]=2$ nilpotent uplift
is tuned to a 4D Minkowski minimum.  The racetrack parameters and axes are
the same as in Fig.~\ref{fig:2}.  The vacuum is the asymptotic finite-$L_5$
valley $L_5M_5\simeq27.76$, $L_4\to\infty$.  Both this valley and the simultaneous
decompactification direction approach $V=0$ from above.}}
\label{fig:nilpotentContour}
\end{figure}

\WY{Consider a nilpotent-multiplet uplift of the racetrack potential used
in the main text,}
\begin{equation}
\WY{
 {\cal K}_{\rm up}={\cal K}_4+X\bar X,\qquad
 W_{\rm up}=W_{\rm rt}+\mu^2X,\qquad X^2=0 .
}
\label{eq:nilpotentUplift}
\end{equation}
\WY{I take $R[X]=2$ and choose $\mu^2\in\mathbb R$ by a phase redefinition
of $X$.  The uplift term respects the $R$ symmetry, and no independent
constant term is present in $W_{\rm up}$.}
{Here the nilpotent field is used only as an infrared parametrization of
the uplift; as noted above, the SUSY-breaking sector itself may have a 5D
origin.}
\WY{At $X=0$, this gives}
\begin{equation}
{
 V_{4D}^{\rm up}
 =V_{4D}+\frac{|\mu^2|^2}{(T+\bar T)^3}.
}
\label{eq:nilpotentPotential}
\end{equation}
\WY{I tune $\mu$ so that the finite-$T$ minimum is Minkowski.  In connecting
to the 3D branch, I use the two sector-resolved gauge switches in
Eq.~\eqref{eq:sectorSwitches}, while retaining the independent uplift term.
The resulting contour is shown in
Fig.~\ref{fig:nilpotentContour}.  For the parameters of Fig.~\ref{fig:2},
the Minkowski conditions give
$M_5t_{\rm M}=27.76$ and
$|\mu^2|^2/M_5^6=7.28\times10^{-28}$.}

\clearpage
\bibliographystyle{apsrev4-1}
\bibliography{ref}

\end{document}